\documentclass[preprint,prb,showpacs,preprintnumbers,amsmath,amssymb]{revtex4-1}

\usepackage{graphicx}

\begin{document}


\title{Temperature evolution of correlation strength in the superconducting state of high-$T_c$ cuprates}

\author{S. Kudo$^1$, T. Yoshida$^1$, S. Ideta$^1$, K. Takashima$^1$, H. Anzai$^2$,
T. Fujita$^2$, Y. Nakashima$^2$, A. Ino$^2$, M. Arita$^3$, H.
Namatame$^3$, M. Taniguchi$^{2,3}$,  K. M.
Kojima$^1$, S. Uchida$^1$, A. Fujimori$^1$}

\affiliation{$^1$Department of Physics, University of Tokyo,
Bunkyo-ku, Tokyo 113-0033, Japan}

\affiliation{$^2$Graduate School of Science, Hiroshima University,
Higashi-Hiroshima 739-8526, Japan}

\affiliation{$^3$Hiroshima Synchrotron Center, Hiroshima
University, Higashi-Hiroshima 739-0046, Japan}

\date{\today}

\begin{abstract}
We have performed an angle-resolved photoemission study of the
nodal quasi-particle spectra of the high-$T_c$ cuprate tri-layer
Bi$_2$Sr$_2$Ca$_2$Cu$_3$O$_{10+\delta}$ ($T_c\sim$ 110 K). The
spectral weight $Z$ of the nodal quasi-particle increases with
decreasing temperature across the $T_c$. Such a temperature
dependence is qualitatively similar to that of the coherence peak
intensity in the anti-nodal region of various high-$T_c$ cuprates
although the nodal spectral weight remains finite and large above
$T_c$. We attribute this observation to the reduction of electron
correlation strength in going from the normal metallic state to
the superconducting state, a characteristic behavior of a
superconductor with strong electron correlation. 
\end{abstract}


\maketitle

The strength of electron correlation in a metal is represented by
the renormalization factor $Z_{k_F} =
|<\Phi_{k_F}(N-1)|a_{k_F}|\Phi_{k_F}(N)>|^2$, where $a_{k_F}$ is
the annihilation operator of electron on the
Fermi surface, $\Phi_{k_F}(N)$ is the ground state, and
$\Phi_{k_F}(N-1)$ is the lowest energy state with a quasi-particle
(QP) of momentum $k_F$. $Z_{k_F}$ gives the weight of the QP
peak in the single-particle spectral function $A(k,\varepsilon)$ at $k=k_F$
 and $\varepsilon$ at the Fermi level ($E_F$),
which can be measured by angle-resolved photoemission spectroscopy
(ARPES)\cite{Hufner}. According to BCS theory, which describes
superconductivity in metals with weak electron correlation, the QP
is fully coherent ($Z_{k_F}$ =1) all over the Fermi surfaces and,
in a $d$-wave BCS superconductor, the QP spectrum on the nodal
point of the Fermi surface remains unchanged without gap opening
and with $Z_{k_F}$=1. One subtle change that has been noticed so
far is the sharpening of the QP width below $T_c$ (ref. \onlinecite{Yamasaki}). 
If superconductivity occurs in the presence of strong
electron correlation, on the other hand, the nodal spectral weight increases in going from
the normal state to the superconducting state as theoretically predicted by
Chou, Lee and Ho \cite{Chou}. In the theory, the
correlated normal state is represented by a projected Fermi-liquid
and the correlated superconducting state by a
resonating valence-bond (RVB) or a projected BCS state. The
increase of the QP spectral weight $Z_{k_F}$ reflects the partial
recovery of the coherence from the highly incoherent projected Fermi-liquid state to in the projected BCS state.

So far, the temperature dependence of the Drude weight in optical conductivity has been investigated for various cuprates \cite{Carbone, Ortolani, Hwang, Bontemps}. In those studies, the low energy electron number defined by the integration of optical conductivity below a certain cut-off frequency gradually  increases with decreasing temperature. Such a behavior is also observed in normal metals such as gold, but the change in the cuprates, e. g., La$_{2-x}$Sr$_x$CuO$_4$ (LSCO), is more than one order of magnitude as large as that in gold. The optical conductivity can be expressed by a two-particle Green's function, and the QP spectral weight which is derived from the one-particle Green function is also expected to show a similar temperature dependence. However, direct test of the temperature dependence of the QP spectral weight has not been investigated in a systematic way. In the present work, we have performed an ARPES study of the tri-layer cuprates superconductor
Bi$_2$Sr$_2$Ca$_2$Cu$_3$O$_{10+\delta}$ (Bi2223) in a wide
temperature range and investigated changes in the spectral weight
of the nodal QP with temperature. The results indeed show a
dramatic increase of the QP spectral weight with decreasing
temperature.

Single crystals of Bi2223 ($T_\mathrm{c}$ = 110 K) were grown by
the traveling solvent floating-zone (TSFZ) method. ARPES
measurements were carried out at BL-9A of Hiroshima Synchrotron
Radiation Center (HiSOR). Incident photons have an energy of
$h\nu$ = 7.56 eV, and measurements were made at $T$ = 11, 90, 120,
and 160 K. A SCIENTA SES-R4000 analyzer was used in the angle mode
with the total energy and momentum resolution of $\sim$ 5 meV and
$\sim$ 0.3$^\circ$, respectively. All the samples were cleaved
\textit{in situ} under an ultrahigh vacuum of 10$^{-11}$ Torr. The
position of the $E_\mathrm{F}$ was calibrated with gold spectra.

The single-particle spectral function $A(\mathbf{k},\varepsilon)$
measured by ARPES is the imaginary part of the single-particle
Green's function $G(\mathbf{k},\varepsilon)\equiv
1/(\varepsilon-\varepsilon_\mathbf{k}-\Sigma(\mathbf{k},\varepsilon))$:
\begin{eqnarray}
A(\mathbf{k},\varepsilon)&\equiv&-\frac{1}{\pi}\mathrm{Im}G\nonumber\\
&=&-\frac{1}{\pi}\frac{\mathrm{Im}\Sigma(\mathbf{k},\varepsilon)}{(\varepsilon-\varepsilon_\mathbf{k}-\mathrm{Re}\Sigma(\mathbf{k},\varepsilon))^2+(\mathrm{Im}\Sigma(\mathbf{k},\varepsilon))^2},\label{A1}
\end{eqnarray}
where $\varepsilon_\mathbf{k}$ is the bare band electron energy with momentum $\mathbf{k}$ and $\Sigma(\mathbf{k},\varepsilon)$ is
the self-energy. $\varepsilon=0$ is chosen at $E_\mathrm{F}$. The pole of
$\mathrm{Re}G(\mathbf{k},\varepsilon)$,
$\varepsilon=\varepsilon_\mathbf{k}^*$ is determined by the
equation$\varepsilon-\varepsilon_\mathbf{k}-\mathrm{Re}\Sigma(\mathbf{k},\varepsilon)=0$ and the residue of the pole
$Z_\mathbf{k}(\varepsilon_\mathbf{k}^*)$
\begin{equation}
Z_\mathbf{k}(\varepsilon_\mathbf{k}^*)\equiv\left(1-\frac{\partial\mathrm{Re}\Sigma(\mathbf{k},\varepsilon)}{\partial\varepsilon}\Bigg|_{\varepsilon=\varepsilon_\mathbf{k}^*}\right)^{-1}\
(<1)
\end{equation}
gives the spectral weight of QP. In the vicinity of
$\varepsilon=\varepsilon_\mathbf{k}^*$, one can expand
$\mathrm{Re}\Sigma(\mathbf{k},\varepsilon)$ as,
\begin{eqnarray}
\mathrm{Re}\Sigma (\mathbf{k},\varepsilon)&\simeq&\mathrm{Re}\Sigma(\mathbf{k},\varepsilon_\mathbf{k}^*)+\frac{\partial\mathrm{Re}\Sigma(\mathbf{k},\varepsilon)}{\partial\varepsilon}\Bigg|_{\varepsilon=\varepsilon_\mathbf{k}^*}(\varepsilon-\varepsilon_\mathbf{k}^*)\nonumber\\
&\simeq&\varepsilon-\varepsilon_\mathbf{k}-\frac{1}{Z_\mathbf{k}(\varepsilon_\mathbf{k}^*)}(\varepsilon-\varepsilon_\mathbf{k}^*).
\end{eqnarray}
In the vicinity of $E_\mathrm{F}$, $\varepsilon_\mathbf{k}^*\sim
v_\mathbf{k}^*(k-k_\mathrm{F})$, where $k$ is taken perpendicular
to the Fermi surface and
$v_\mathbf{k_F}^*(\equiv|\nabla\varepsilon_\mathbf{k_F}^*|)$ is the
Fermi velocity. Then, the momentum distribution curve (MDC) at
$E_\mathrm{F}$ is given by
\begin{equation}
A(\mathbf{k},0)\simeq-\frac{Z_\mathbf{k_F}(0)/v_\mathbf{k_F}^*}{\pi}\frac{Z_\mathbf{k_F}(0)\mathrm{Im}\Sigma(\mathbf{k},0)/v_\mathbf{k_F}^*}{(k-k_\mathrm{F})^2+(Z_\mathbf{k_F}(0)\mathrm{Im}\Sigma(\mathbf{k},0)/v_\mathbf{k_F}^*)^2}.
\end{equation}
This MDC is a Lorentzian, if the self-energy
$\Sigma(\mathbf{k},\varepsilon)$ is not strongly dependent on $k$
perpendicular to the Fermi surface. The QP weight is, therefore, given by
\begin{equation}
Z_{k_F}(0)=\int_{-\infty}^\infty
A(\mathbf{k},0)d\mathbf{k}\times v_\mathbf{k_F}^*.\label{Z}
\end{equation}

\begin{figure}
\includegraphics[width=16cm]{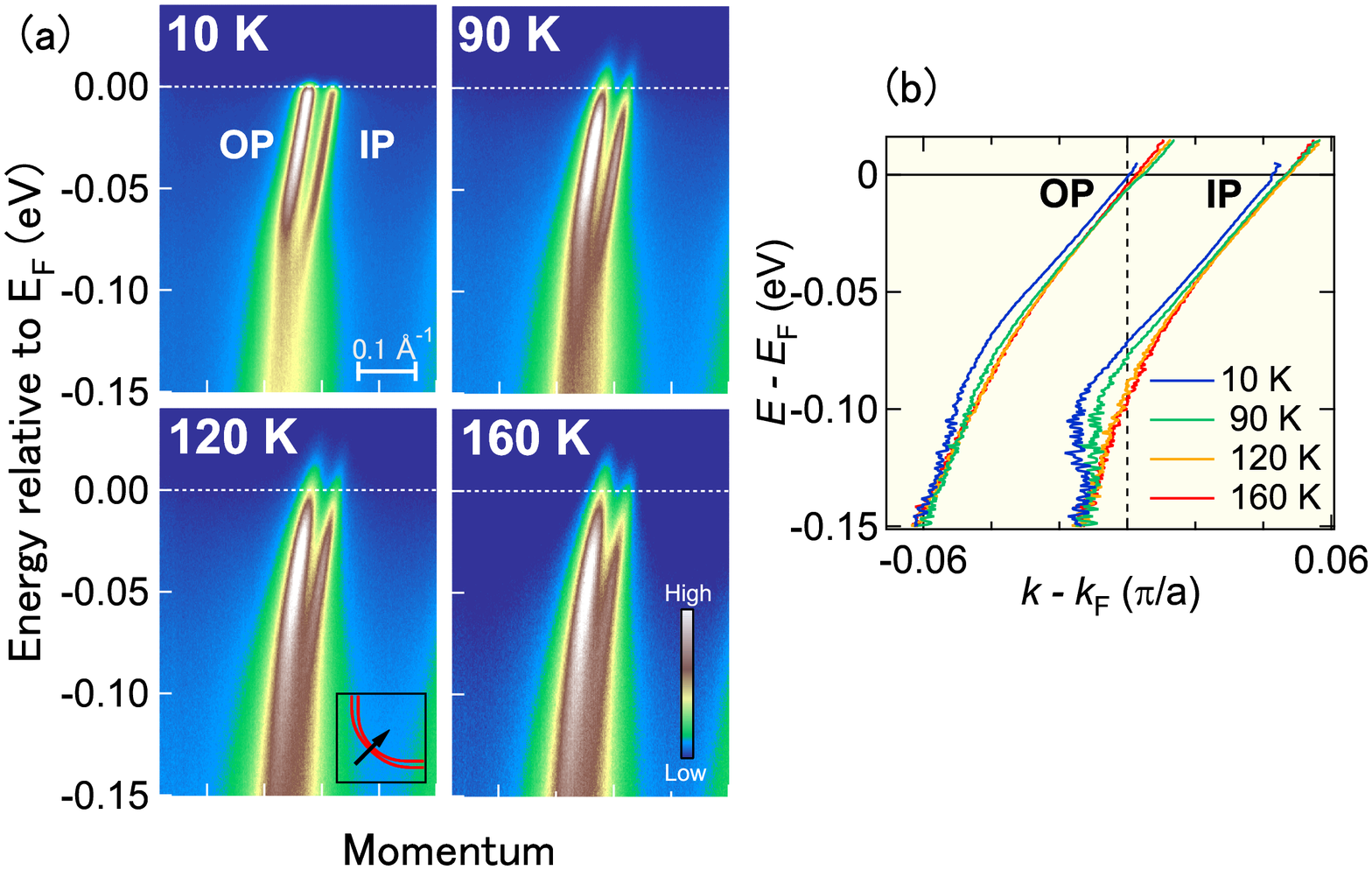}
\caption{\label{Ek}(Color online) ARPES spectra of Bi2223 in the
nodal direction at various temperatures. (a) Intensity plots in
energy-momentum space along the nodal direction (see the inset to
the 120 K data). OP and IP denote the outer and inner
quasi-particle bands, respectively. (b) Quasi-particle (QP) band
dispersions for the IP and OP bands deduced from the MDC peak
positions.}
\end{figure}

Figure \ref{Ek}(a) shows plots of spectral intensities along the nodal $\mathbf{k}=(0,0)-(\pi,\pi)$ cut. There are two bands corresponding to the inner CuO$_2$
plane (IP) and outer CuO$_2$ plane (OP) of the tri-layer cuprate
\cite{Ideta}. 

 In panel (b), the QP dispersions are traced by the peak positions of the MDC's.
One can clearly see kinks in the QP dispersions which show
temperature dependences. From these spectra, we shall deduce the
temperature dependence of the QP spectral weight $Z_{k_F}$ in the
nodal direction.

\begin{figure}
\includegraphics[width=16cm]{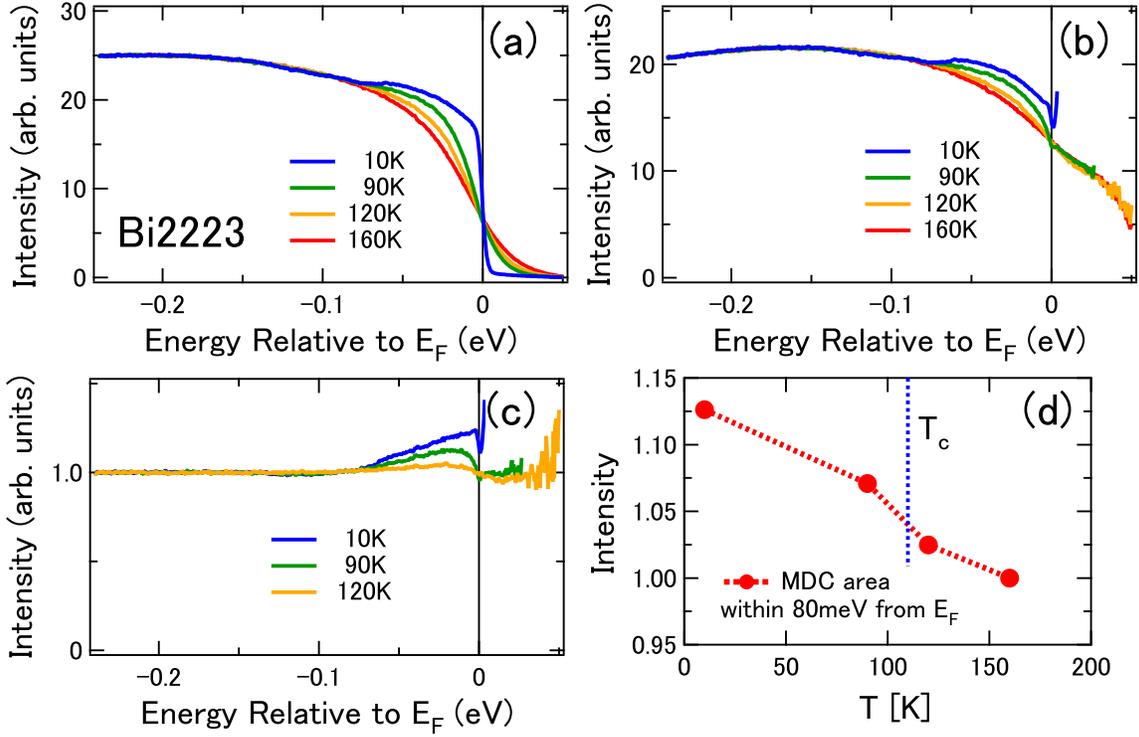}
\caption{\label{IMDC}(Color online) ARPES spectra of Bi2223
integrated along the nodal direction including both the IP and OP
bands. (a) Raw data. (b) Spectra divided by the Fermi-Dirac
function. (c) Spectra in panel (b) divided by the spectrum at 160
K. (d) Temperature dependence of the spectral weight intensity
within 80 meV from $E_F$ (normalized to the 160 K data).}
\end{figure}

First, we look into the temperature dependence of the MDC area,
namely, the momentum-integrated ARPES spectra along the nodal
direction as shown in Fig. \ref{IMDC}(a). Here, the intensity has
been normalized to the intensity at high binding energies $>$ 0.2
eV. Figure \ref{IMDC} (b) shows the same data divided by the
Fermi-Dirac (FD) function (The gap opens in the 10 K data because
the cut was slightly off nodal due to small misalignment). In order to emphasize changes induced by superconductivity, the spectra in panel (b) have been divided by the 160 K spectrum
and are shown in panel (c). Thus obtained spectra clearly indicate
that the intensity within 80 meV of $E_F$ increases with
decreasing temperature. The integrated intensities within
$\sim$80meV are plotted as a function of temperature in panel (d).
Note that the integrated spectral intensities include signals from both the IP
and OP bands.

\begin{figure}
\includegraphics[width=16cm]{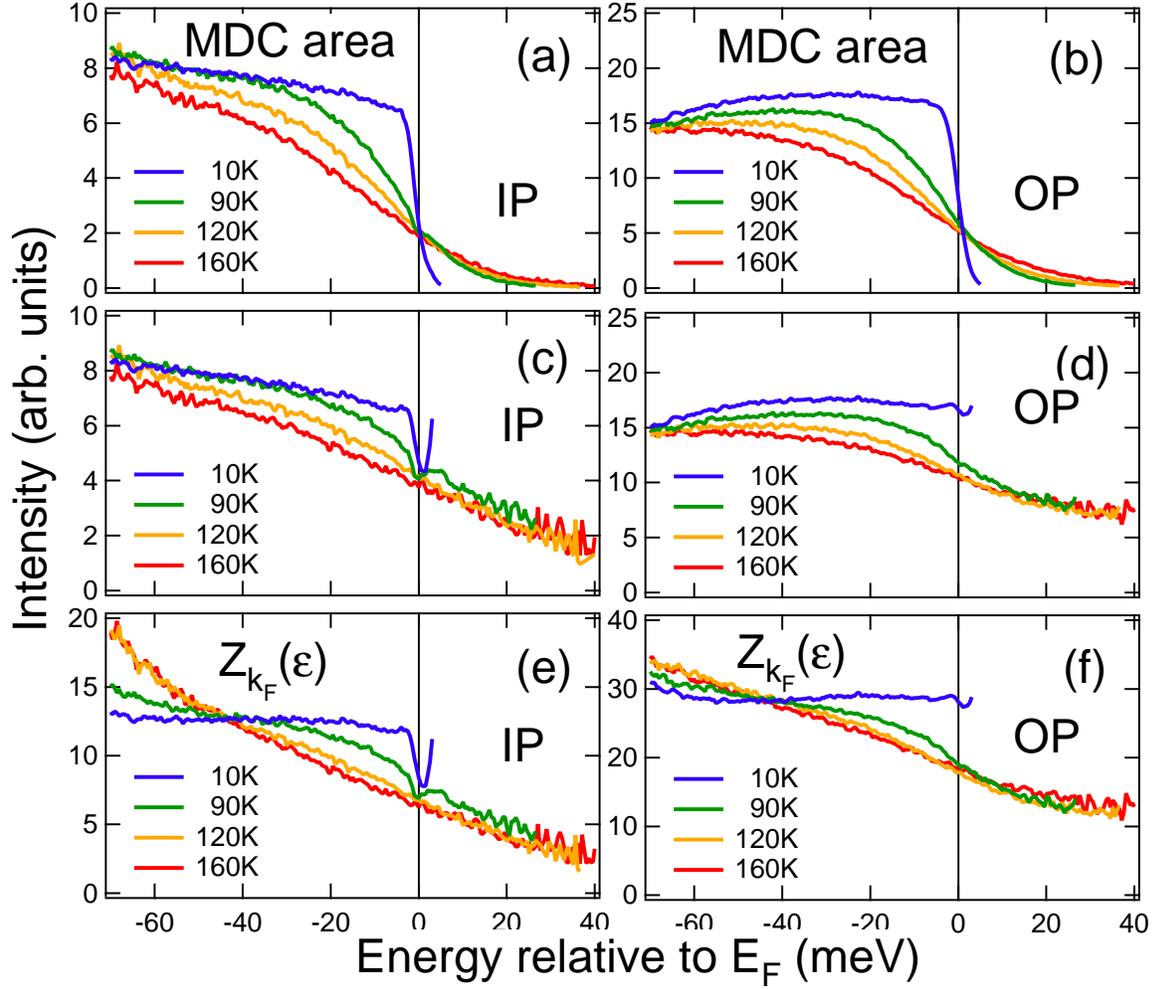}
\caption{\label{Z_MDC}(Color online) Spectral weight
$Z_{k_F}(\epsilon)$ for the IP and OP bands of Bi2223 deduced from
the MDC peak area and the QP velocity $v_k^*(\epsilon)$. In order
to extract the MDC area for the IP and OP bands separately, MDC
data in Fig.\ref{Ek} were fitted to two Lorentzians. (a)(b) MDC
area with the same intensity normalization as Fig. \ref{IMDC}.
(c)(d) MDC area divided by Fermi-Dirac function convoluted with a
Gaussian. (e)(f) $Z_{k_F}(\epsilon)=\int_{-\infty}^\infty
A(\mathbf{k},0)d\mathbf{k}\times v_k^*(\epsilon)$, where
$v_k^*(\epsilon)$ has been deduced from the slope of the QP
dispersions shown in Fig.\ref{Ek}(b). The dip at $E_F$ in the
10 K data, particularly for the IP band, arises because the cut direction was slightly off node due to small misalignment.}
\end{figure}

In order to estimate the QP spectral weight
$Z_{k_F}(0)$ at $E_F$ using Eq. (\ref{Z}), signals from the IP and
OP bands have to be separated. For that purpose, the MDC at each
energy has been fitted to two Lorentzians. The Fermi velocity at
$E_F$, $v_F^*$, has been obtained from the slope of the QP
dispersion shown in Fig.1(b). Here, we extend Eq. (\ref{Z}), which is defined at
$E_F$, to finite energies as:
$Z_{kF}(\varepsilon)f(\varepsilon)\propto \int_{-\infty}^\infty
A(\mathbf{k},\varepsilon)d\mathbf{k} \times v_k(\varepsilon)$. The
momentum-integrated spectrum for each of the IP and OP bands at
various temperatures is plotted in Figs.\ref{Z_MDC}(a) and (b). Since the observed MDC area is $f(\varepsilon)\int_{-\infty}^\infty
A(\mathbf{k},\varepsilon)d\mathbf{k}$, where $f(\varepsilon)$ is
the FD distribution function, we have divided the spectra by the FD
function as plotted in Figs.\ref{Z_MDC}(c) and (d). The
$Z_{k_F}(\varepsilon)$ spectra thus deduced using the finite energy version of Eq. (\ref{Z})
plotted in Figs. \ref{Z_MDC}(e) and (f).

$Z_{k_F}(\varepsilon)$ at $T$=10 K is nearly constant in the displayed energy range near $E_F$ as expected for a Fermi liquid or a QP at the node of a $d$-wave BCS
superconductor, while $Z(\varepsilon)$ at high temperatures (120 K
and 160 K ) decrease towards $E_F$ and above it. Also, the $Z_{k_F}(0)$ value at $E_F$ itself decreases with increasing temperature.  This indicates that the nodal spectrum is highly coherent well below $T_c$, but gradually becomes incoherent with increasing
temperature. The degree of deviation from constant
$Z_{k_F}(\varepsilon)$ or the loss of coherence is a little stronger
for the IP band than the OP band, probably due to the smaller hole
concentration and hence the stronger electron correlation in the
IP band$\cite{Ideta}$.

Although less accurate, the spectral weight $Z_{k_F}$ can also be derived from
energy distribution curves (EDC's), under the assumption that the normal state is a Fermi liquid (although the high-$T_c$ cuprates near optimum doping is considered to be a marginal Fermi liquid \cite{Varma}). For a Fermi liquid,
$\Sigma(\mathbf{k},\varepsilon)$ can be expanded in the vicinity
of $E_\mathrm{F}$ as
\begin{equation}
\Sigma(\mathbf{k},\varepsilon)\simeq-\alpha_\mathbf{k}\varepsilon-i\beta_\mathbf{k}\varepsilon^2\
(\alpha_\mathbf{k},\ \beta_\mathbf{k}>0).\label{Sigma}
\end{equation}
From Eqs. (\ref{A1}) and (\ref{Sigma}), therefore, one obtains
\begin{eqnarray}
A(\mathbf{k},\varepsilon)&\simeq&\frac{1}{\pi}\frac{\beta_\mathbf{k}\varepsilon^2}{(\varepsilon-\varepsilon_\mathbf{k}+\alpha_\mathbf{k}\varepsilon)^2+(\beta_\mathbf{k}\varepsilon^2)^2}\nonumber\\
&=&\frac{Z_\mathbf{k}}{\pi}\frac{Z_\mathbf{k}\beta_\mathbf{k}\varepsilon^2}{(\varepsilon-Z_\mathbf{k}\varepsilon_\mathbf{k}^0)^2+(Z_\mathbf{k}\beta_\mathbf{k}\varepsilon^2)^2},\label{A3}
\end{eqnarray}
where $Z_\mathbf{k}\equiv(1+\alpha_\mathbf{k})^{-1}\ (<1)$.
When $\mathbf{k}$ is not on the Fermi surface
($\mathbf{k}\neq\mathbf{k}_\mathrm{F}$), in the vicinity of
$E_\mathrm{F}$,
\begin{equation}
A(\mathbf{k},\varepsilon)\simeq\frac{\beta_\mathbf{k}}{\pi}\frac{\varepsilon^2}{\varepsilon_\mathbf{k}^2}\propto-\mathrm{Im}\Sigma(\mathbf{k},\varepsilon),
\end{equation}
and, therefore, there is no spectral weight at $E_\mathrm{F}$. As
$\mathbf{k}$ approaches the Fermi surface, the QP peak width
becomes narrow, and when $\mathbf{k}$ is on the Fermi surface
($\mathbf{k}=\mathbf{k}_\mathrm{F}$), the QP peak becomes a
$\delta$-function $Z_\mathbf{k}\delta(\varepsilon)$:
\begin{equation}
A(\mathbf{k}_\mathrm{F},\varepsilon)\simeq Z_k\delta(\varepsilon)+
 \frac{Z_{\mathbf{k}_\mathrm{F}}}{\pi}\frac{Z_{\mathbf{k}_\mathrm{F}}\beta_{\mathbf{k}_\mathrm{F}}}{1+Z_{\mathbf{k}_\mathrm{F}}^2\beta_{\mathbf{k}_\mathrm{F}}^2\varepsilon^2}\label{A4}.
\end{equation}
That is, the line shape of the EDC at
$\mathbf{k}=\mathbf{k}_\mathrm{F}$ is a $\delta$-function
superposed on top of a broad Lorentzian-like background both
centered at $E_F$ \cite{Allen}. Here, the $\delta$-function is broadned into a Lorentzian  due to impurity scattering. Experimentally, therefore, we expect
to observe overlapping two Lorentians with narrow and broad
widths. 

\begin{figure}
\includegraphics[width=16cm]{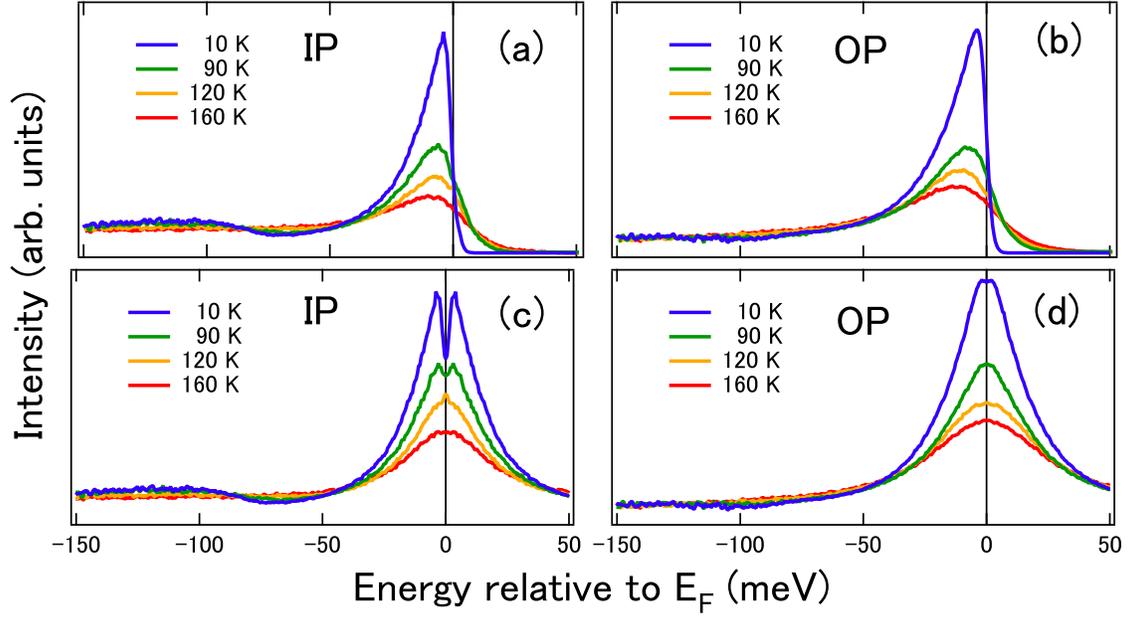}
\caption{\label{Z_kF}(Color online) EDC's of Bi2223 at $k_F$ on
the node. (a)(b) Temperature dependence of the EDCs at $k_F$.
(c)(d) Symmetrized EDCs at $k_F$. The area of the Lorentzian
centered at $E_F$ gives $Z(0)$.}
\end{figure}

Figure \ref{Z_kF} shows EDC's at
$k=k_\mathrm{F}$ [(a),(b)] and their symmetrized spectra
[(c),(d)]. As shown in panels (c) and (d)  for the IP and OP bands, the line shape of each
symmetrized spectrum shows a Lorentzian-like central peak. (The splitting of  the
symmetrized 10 K data at $E_F$, particularly in the IP data, arises from small misalignment of the nodal direction.) Again the increase of the spectral weight
of the central peak is seen with decreasing temperature. According to Eq.
(12), a broad Lorenztian-like background is expected in addition to the relatively sharp Lorenzian QP peak of spectral weight
$Z_{k_F}(0)$. However, because such a background feature appears
negligibly small, we have estimated $Z_{k_F}(0)$ from
the area of the symmetrized EDC's within 80 meV of $E_F$.

\begin{figure}
\includegraphics[width=16cm]{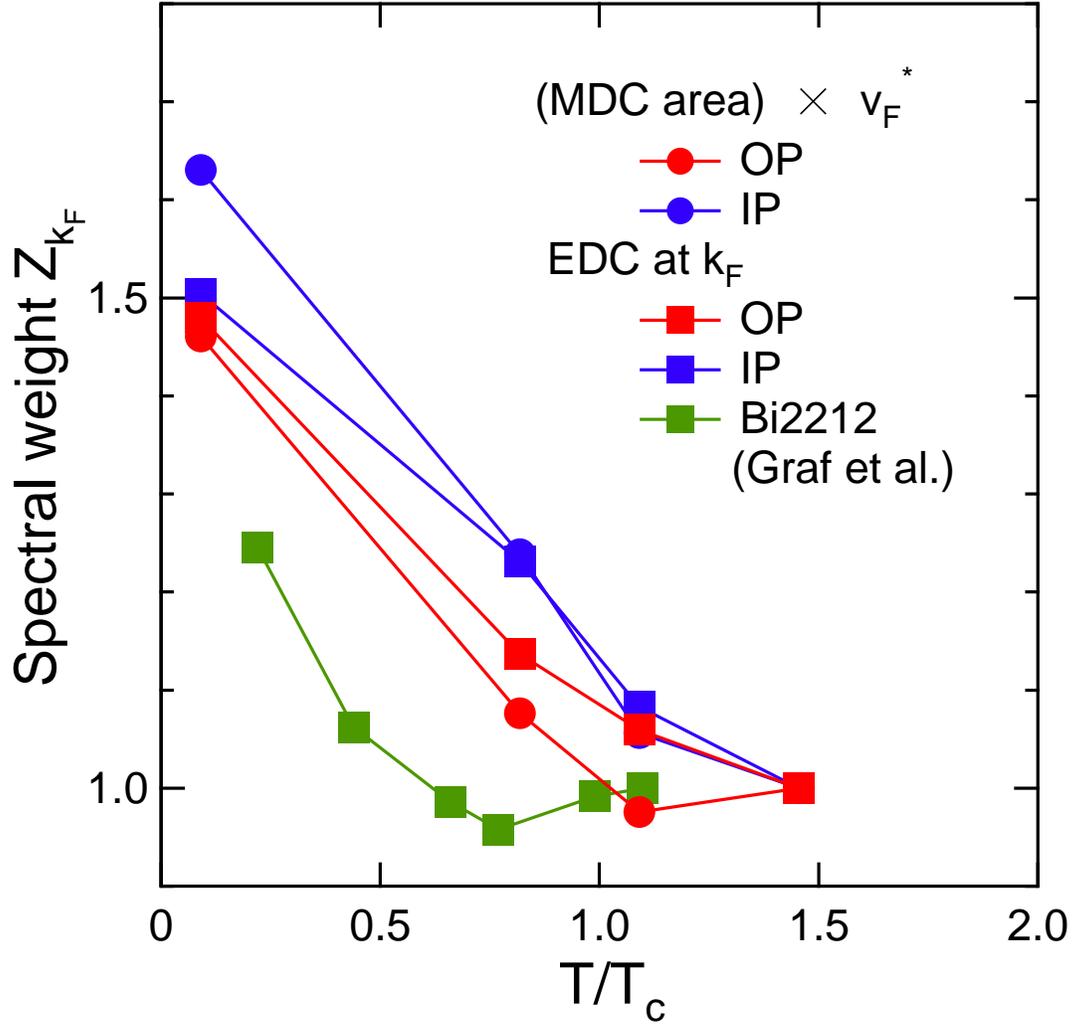}
\caption{\label{Z_T}(Color online) Spectral weight at $E_F$,
$Z_{k_F}(0)$, for the OP and IP bands of Bi2223 derived from MDC's and EDC's (under the assumption of a Fermi liquid) plotted as functions of temperature. $Z_{k_F}(0)$ for Bi2212 ($T_c$=91 K) deduced from the EDC's in ref. \onlinecite{Graf} is also plotted.}
\end{figure}

The temperature dependence of the nodal spectral weight
$Z_{k_F}(0)$ for the IP and OP bands obtained from the MDC and
EDC analysis are summarized in Fig. \ref{Z_T}. The values of
$Z_{k_F}(0)$ at various temperatures have been normalized to the
values at 160 K. The temperature dependence of $Z_{k_F}$ obtained
by the MDS's and EDC's fall almost on the same curve. The results show
a clear increase of $Z_{k_F}$ by as much as $\sim$50\% in going
from 160 K to 10 K. The $Z_{k_F}$ of the IP band increases a little
more rapidly than that of the OP band with decreasing temperature.
Because the IP and OP bands are underdoped and overdoped,
respectively, the result indicates that the underdoped CuO$_2$
plane loses coherence faster than the overdoped CuO$_2$ plane with
temperature. Here, we have also plotted the $Z_{k_F}(0)$ of Bi2212 in ref. \onlinecite{Graf}.
The Bi2212 data also show a similar temperature dependence to that of Bi2223. The smaller increase in the $Z_{k_F}$ of Bi2212 with decreasing temperature than that in Bi2223 may reflect the smaller superconducting order parameter.   

A similar increase of the low energy spectral weight has been observed in optical conductivity \cite{Carbone,Bontemps,Ortolani,Hwang}, indicating an increase of the coherence in the superconducting state compared to the normal state. In the underdoped region, the Drude weight of the high-$T_c$ cuprates shows a strong temperature dependence compared to that in the overdoped region, which can be interpreted as an electron correlation and/or pseudogap effects \cite{Santander, Bontemps}. Particularly in the underdoped region, the low energy spectral weight, i.e., the electron kinetic energy, strongly increases below $T_c$ \cite{Molegraaf}. Since the contribution from the kinetic energy to the condensation energy is significant compared to the conventional superconductors, kinetic energy driven superconductivity has been proposed \cite{Maier, Norman_T}. The increase in the spectral weight below $T_c$ is up to 50\%, which is much larger than the change in the Drude weight in various high-$T_c$ cuprates. Such a strong increase of the spectral weight, and hence the increase in the kinetic energy may be one of the reasons of the very high $T_c$ in the tri-layer cuprates.

As shown in Figs. \ref{Z_MDC}(e) and \ref{Z_MDC}(f), $Z(\varepsilon)$ below $T_c$ shows a nearly energy-independent Fermi-liquid-like behavior, which is in contrast to strongly energy-dependent incoherent behavior above $T_c$. Particularly, the $Z(\varepsilon)$ of IP monotonically decrease with energy, indicating a strongly incoherent nature of the spectral weight. In the previous photoemission studies, the coherence temperature $T_{coh}$ has been deduced from the line shape of the spectra \cite{Kaminski, Hashimoto} and has been found to increase monotonically with hole concentration, consistent with the prediction of the $t$-$J$ model. In the present result, even though the nodal spectral weight is most coherent on the Fermi surface, the spectra become highly incoherent at high temperatures, also consistent with the RVB picture based on the $t$-$J$ model.

In conclusion, we have performed a temperature-dependent
angle-resolved photoemission spectroscopy study of the optimally
doped tri-layer high-$T_\mathrm{c}$ cuprates Bi2223 to investigate
the temperature dependence of spectral weight $Z_{k_F}$ in the
nodal direction. In contrast to what is expected from BCS theory, all the results
show an increase of spectral weight $Z_{k_F}$ with decreasing
temperature below $T_c$, consistent with the theoretical prediction on correlated superconductors \cite{Chou} and suggests a transition from the relatively
incoherent metal to the relatively coherent superconductor across $T_c$. The result indicates not only the change in the coherence at $T_c$ but also the rapid evolution of the coherence with decreasing
temperature below $T_c$. 

We would like to thank fruitful and enlightening discussion with T.K. Lee, C.M. Ho, and C.P. Chou. ARPES experiments were carried out at HiSOR (Proposal No. 07-A-10, No. 08-A-35).

\bibliography{SW}

\end{document}